\documentclass[letterpaper,twocolumn,10pt]{article}
\usepackage{usenix2019_v3}

\usepackage{tikz}
\usepackage{filecontents}

\usepackage[normalem]{ulem}
\usepackage{epsfig,endnotes}
\usepackage{amsmath}
\usepackage{amssymb}
\usepackage{datetime}
\usepackage{booktabs}
\usepackage[linesnumbered,ruled]{algorithm2e}
\usepackage{pifont}
\usepackage{url}
\usepackage{cite}

\usepackage{paralist}
\usepackage{marvosym}
\usepackage{multirow,array}
\usepackage{graphicx}
\usepackage{soul}
\usepackage{todonotes}

\newcommand{\showComments}{yes}

\newcommand{\note}[2]{
    \ifthenelse{\equal{\showComments}{yes}}{\textcolor{#1}{#2}}{}
}

\begin{document}
%
% The "title" command has an optional parameter, allowing the author to define a "short title" to be used in page headers.
\title{Kernel/User-level Collaborative Persistent Memory File System with Efficiency and Protection}
\date{}

%for single author (just remove % characters)
\author{
{\rm Youmin Chen}
\and
{\rm Youyou Lu}
\and
{\rm Bohong Zhu}
\and
{\rm Jiwu Shu}
\and
{\rm Tsinghua University}
% copy the following lines to add more authors
% \and
% {\rm Name}\\
%Name Institution
} % end author

%
% This command processes the author and affiliation and title information and builds
% the first part of the formatted document.
\maketitle

%
% The abstract is a short summary of the work to be presented in the article.
\begin{abstract}
\vspace{-0.05in}
Emerging high performance non-volatile memories recall the importance of
\textit{efficient} file system design. To avoid the virtual file system (VFS)
and \texttt{syscall} overhead as in these kernel-based file systems,
recent works deploy file systems directly in user level.
Unfortunately, a user level file system can easily be corrupted by a buggy program
with misused pointers, and is hard to scale on multi-core platforms which incorporates a
centralized coordination service.

In this paper, we propose KucoFS, a \underline{K}ernel and \underline{u}ser-level
\underline{co}llaborative file system.
It consists of two parts: a user-level library with direct-access interfaces,
and a kernel thread, which performs metadata updates and enforces
write protection by toggling the permission bits in the page table.
Hence, KucoFS achieves both \textsf{direct-access} of user-level
designs and fine-grained \textsf{write protection} of kernel-level ones.
We further explore its scalability to multicores:
For metadata scalability, KucoFS rebalances the pathname resolution overhead between the
kernel and userspace, by adopting the index offloading technique.
For data access efficiency, it coordinates the data allocation between kernel and userspace,
and uses range-lock write and lock-free read to improve concurrency.
% the indexing lookup
% from the kernel to the userspace, and batches the metadata operation logs.
% For data access efficiency, KucoFS preallocates data blocks from kernel to userspace
% and introduces range-lock for data writes, while making reads lock-free.
Experiments on Optane DC persistent memory show that KucoFS significantly
outperforms existing file systems and shows better scalability.

\iffalse
Existing operating systems abstract storage components via a unified virtual file
system and a group of \emph{syscall} based interfaces. Such classical designing
is inherently suitable for traditional storage devices, but dramatically limits
the performance of emerging hardwares like multi-core
CPUs and non-volatile memories. In this paper, we revisit the file system
architecture by proposing KucoFS, which decouples the file system into two parts:
A user-level library to enable \emph{direct file access},
and a dedicated kernel thread to manage the control-plane operations.
% Different from existing user-level file systems,
KucoFS doesn't compromise the properties that a kernel file system has and
is capable of providing both safe, efficient and scalable data storage.
It achieves these goals by leveraging the root privileges of a kernel thread to
enforce fine-grained protection, and carefully balancing the loads between kernel
and user space. Our experiments show that
KucoFS achieves the highest performance among all the evaluated file systems and
can always be scalable despite the skewness of the workloads.
\fi
\end{abstract}

\section{Introduction}

Emerging byte-addressable non-volatile memories (NVMs), such as
PCM~\cite{isca09pcmlee, isca09pcmqureshi, isca09pcmzhou},
ReRAM~\cite{Baek2004HighlySN}, and the
recently released Intel Optane DC persistent memory~\cite{3dxpoint},
provide performance
comparable to DRAM and data persistence similar to disks.
Such high-performance hardware recalls the importance of redesigning
\emph{efficient} file systems.
The efficiency refers to not only the lightweight software overhead of the
file system itself, but also its scalability to multicores that is able
to exploit the hardware performance of non-volatile memories.

File systems have long been part of an operating system, and are placed
in the kernel level to provide data protection from arbitrary user writes.
System calls (\texttt{syscall}s) are used for the communication between
the kernel and userspace.
In the kernel, the virtual file system (VFS) is an abstraction layer
that hides concrete file system designs to provide uniform accesses.
However, both \texttt{syscall} and \texttt{VFS} incur non-negligible overhead
in file systems for NVMs.
Our evaluation on NOVA~\cite{fast16nova} shows that, even the highly scalable and efficient NVM-aware
file system still suffers great overhead in the VFS layer and fails to
scale on some file operations (e.g., \texttt{creat/unlink}).
For \texttt{syscall}, the context switch overhead occupies
up to 34\% of the file system accessing time, even without
counting the effects of TLB and CPU cache misses on the following execution.

Recent works like Strata~\cite{sosp17strata} and Aerie~\cite{eurosys14aerie}
propose to design NVM file systems in the user level.
By bypassing the operating system, they exploit the
benefits of \emph{direct access}. However, since the NVM space
is exported to applications' address space, a programmer can easily corrupt the file system image
by misusing pointers, which accidently point to the NVM space.
Moreover, these file systems adopt a trusted, but centralized component
to coordinate the critical updates, which inevitably restricts their scalability to multi-cores.

It is difficult to achieve both performance efficiency and write protection simultaneously,
as long as the VFS and kernel/user-space architecture remain unchanged.
% We believe that a conservative approach will not solve the above problems
% fundamentally so long as the VFS and the Linux kernel architecture remain unchanged.
% Inspired by existing works that achieve \emph{direct access} by implementing the
% file systems in user space~\cite{eurosys14aerie, sosp17strata},
In this paper, we revisit the file system
architecture and propose a \underline{K}ernel and \underline{u}ser-level
\underline{co}llaborative File System named KucoFS. Unlike existing user-level file systems,
KucoFS enables user-level direct-access while ensuring write protection that
a kernel file system provides.
KucoFS decouples the file system into a kernel thread (a.k.a., \emph{master}) and
a user space library (a.k.a., \emph{Ulib}).
Programs are capable of directly reading/writing file data in user level
by linking with \emph{Ulib}, while the \emph{master} is dedicated
% to coordinate the concurrent updates and guarantee the integrity of the file system metadata.
to updating metadata on behalf of the applications, as well as guaranteeing the
integrity of file data.
KucoFS prevents a buggy program from corrupting the file system
by exporting the NVM space to user level in \textbf{read-only} mode.
In this way, the read operations still can be conducted in user space.
To serve write operations without compromising the direct-access feature,
the \emph{master} carefully manipulates the page table to make
the related data pages writable beforehand,  and
read-only again once the operation completes, retaining the protection feature.
%This is achievable since the \emph{master} being in kernel space can easily modify the
%permission bits in the page table leveraging its privileged properties.

We further explore the multicore scalability from the following aspects:
\ding{202} \ul{\emph{Metadata Scalability.}}
Like existing user-level file systems, KucoFS introduces a centralized \emph{master},
despite the different insight behind such architecture.
As a result, the \emph{master} in KucoFS is still the bottleneck
when the number of served programs increases.
We introduce \emph{index offloading} to migrate the pathname resolution overhead to userspace,
and use \emph{batching-based logging} to amortize the metadata persistence overhead.
\ding{203} \ul{\emph{Write Protocol.}} To write file data, \emph{Ulib} needs
to interact with the \emph{master} both before and after the operation to enforce write protection.
This can not only further increase the pressure on the \emph{master}, but also
lead to increased latency.
We propose an efficient write protocol to reduce the number of interactions
between \emph{Ulib} and \emph{master} when writing a file. It achieves this by
lazily reserving free data pages from the \emph{master}, and
coordinating the concurrent write operations directly in user space with a range lock.
\ding{204} \ul{\emph{Read-Write Conflicts.}} \emph{Ulib} is likely to read inconsistent data
when it directly accesses the file system without any coordination, while
\emph{master}-involved reading reduces the benefits of direct-access.
We propose \emph{lock-free fast read} to deal with read-write conflicts.
By carefully checking the status of metadata, \emph{Ulib} is able to consistently read file
data without any interacting to the \emph{master}, despite the concurrent writers.

\section{Background}
\label{sec:moti}

\subsection{Kernel File Systems}
Implementing an NVM file system in Linux kernel faces two types of
unavoidable costs, which are the \emph{syscall} overhead and the heavy-weight
software stack in VFS. We investigate the overhead of them
by analyzing NOVA~\cite{fast16nova}, a well-known highly scalable and efficient
NVM-based file system. Our experimental platform is described in Section~\ref{subsec:testbed}.
%We use DRAM to emulate NVM (with read latencies of 300~ns and write bandwidth of
%7.8~GB/s~\cite{fast16nova,sosp17strata}).

\begin{figure}
	\centering
	\includegraphics[width=1\linewidth]{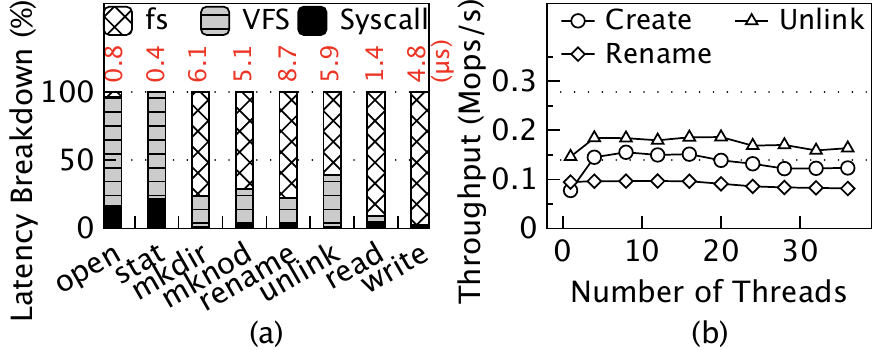}
	\vspace{-0.3in}
	\caption{Analysis of OS-part Overhead with NOVA.}
	\label{fig:moti}
	\vspace{-0.25in}
\end{figure}

\noindent
\textbf{Syscall Overhead.} We analyze the syscall overhead by collecting the
context-switch latency of common file system operations (Each operation is repeated
over 1 million files or directories with a single thread). The results are shown in
Figure~\ref{fig:moti}~(a).
We observe that the context-switch latency takes up to 21\% of the total
execution time, and this ratio is especially large for read-oriented operations (e.g.,
\texttt{stat/open}). Note that the context-switch latency we captured only includes
the direct parts. The indirect costs
(e.g., cache pollution) can further affect the efficiency of
a program~\cite{osdi10flexsc}.

\noindent
\textbf{Inefficiency of VFS.}
In existing Linux kernel, VFS improves the performance of storage devices (e.g., HDD/SSD)
by maintaining page cache in DRAM, but such caching mechanism is not always
effective for NVMs since they have very close access latency. Therefore,
a number of NVM-aware file systems choose to bypass them directly
\cite{fast16nova, eurosys14pmfs, sosp09bpfs, eurosys16hinfs,tos18hinfs,hotstorage18byvfs,cf18spfs}.
However, we find that the remaining software stack in VFS is still too heavyweight:
Our experiments show that NOVA has to spend an average of 34\% of the execution time in VFS layer
(in Figure~\ref{fig:moti}~(a)).
In addition, VFS synchronizes the concurrent syscalls by using the coarse-grained lock, which
limits the scalability. As shown in Figure~\ref{fig:moti}~(b),
to create/rename/delete files in the same folder, VFS directly locks the parent directory, so 
their throughput is unchanged despite the increasing number of client threads.

To sum up, the unified abstraction in VFS and syscall interfaces do provide
a safe and convenient way for programmers, but at the same time,
such classical design concept also restricts us from reconstructing the
file system stack.

\subsection{User Space File Systems}
% To directly access NVM devices without trapping into the kernel,
A group of file systems reduces the OS-part overhead by enabling user-level programs
to directly access NVM devices without trapping into the kernel~\cite{sosp17strata, eurosys14aerie}.
However, they fail to provide the following important properties:
% Kannan et al.~\cite{fast18devfs} classifies them as ``\emph{hybrid file systems
% with trusted server}''.

\noindent
\textbf{Write Protection.}
Both Aerie~\cite{eurosys14aerie} and Strata~\cite{sosp17strata} rely on hardware
virtualization capabilities in modern server systems (i.e., MMU) to enforce coarse-grained
protection: they specify access rights for each application to contiguous subsets of NVM space, so
as to prevent other malicious processes from corrupting the file system image.
However, mapping (a subset of) the file system image to applications' address space and granting them
with write access right is still dangerous, despite that they adopt a third-party service to manage the metadata:
Applications can access Aerie by directly updating the file data in place.
Strata allows user-level programs to directly update
the per-process operation log and the DRAM cache (including both metadata and data).
As a result, a buggy program can easily corrupts the file
system image by misusing pointers which accidentally point to the NVM
space~\cite{asplos11nvheaps, sosp17nova-fortis}, and
the real-world evidence shows that such accidents are really common.

%exposes the NVM device to user-mode programs for
%direct access. It introduces a third-party trusted service (a separated
%user space process) to manage the file system metadata, and a
%distributed locking service for safe sharing between different applications.
%Aerie modifies Linux kernel to only provide coarse-grained memory allocation and
%protection. It also implements a group of
%self-defined interface by relaxing POSIX semantics for higher performance.

\noindent
\textbf{Multicore Scalability.}
Aerie relies on a trusted file system service (TFS, a separate user-level process) to
ensure the integrity of metadata updates and coordinate the concurrent accesses with a
distributed lock service. Such centralized service easily becomes the bottleneck
when the number of concurrent applications increases. Strata,
in contrast, enables applications to update file data by appending their
modifications directly to the per-process log
without the involvement of a third-party service. However, Strata
requires background threads (KernFS) to asynchronously digest the log entries (including both data and metadata)
to the storage devices. If an application completely uses up its log, it must wait for an
in-progress digest to complete before it can reclaim log space. Consequently, the number of digestion threads determines
Strata's overall performance.
Similar to Aerie, Strata also relies on the KernFS for concurrency control, this indicates that
the application needs to interact with the KernFS each time it accesses a new file. Besides,
both of them access the third-party service via socket-based RPCs, which again introduce context-switch overhead,
thus reducing the benefits of direct access.

\section{System Goals}
\label{sec:goal}
In this section, we discuss the designing goals and non-goals and clarify the trade-offs we made when building KucoFS.

\noindent
\textbf{Direct-access and data protection.}
The key design aspect of KucoFS lies in decoupling the functionality of a file system into two parts,
so as to achieve the respective advantages of \emph{direct-access} of user-level file systems,
and \emph{write protection} of kernel-based ones. Note that
KucoFS mainly target at enforcing write protection over buggy programs, and the immunity to malicious
attacks is out of the scope of this paper. Nevertheless, KucoFS is still robust to them in most cases by
using checksum and lease (Section~\ref{subsec:write}).

\noindent
\textbf{Scalability.} KucoFS should work well on a multi-core platform, so as to take full advantages
of the internal parallelism of persistent memory. This drives us to design a scalable \emph{master}
service when it is accessed by concurrent applications. Besides, more efficient concurrency control is
also required to deal with concurrent accesses and read-write conflicts.

\noindent
\textbf{Atomicity and Consistency,} as is required by most existing applications~\cite{osdi14remzi}.
two aspects need to be taken into consideration: 1) KucoFS should always remain consistent
even after the system crashes abnormally, which requires us to carefully design failure atomic update protocols.
2) The readers always see consistent data/metadata when other programs are concurrently
updating files or directories.

\noindent
\textbf{Compatible APIs.} KucoFS should be backward compatible with kernel file system APIs, so that existing applications
can use KucoFS without modifying the source code.

KucoFS makes a few tradeoffs that deviate from standard POSIX semantics, but without
restricting its applicability in real-world applications.
1) KucoFS implements per-user directory trees (i.e., the programs within the same user share a ``private''
root node), instead of a global tree, so as to enforce read protection (Section~\ref{subsec:read}).
%This is not like FlatFS
%in Aerie~\cite{eurosys14aerie}, which totally abandons the POSIX semantics with a flatted key-value
%store interface.
2) We don't provide a explicit way for sharing data between different users as it is not the common case (several feasible
approaches have been proposed in Section~\ref{subsec:read}).
3) Some minor properties are not implemented in KucoFS (e.g., \emph{atime}, etc.).

\section{Design}
\label{sec:design}

We designed KucoFS with the main goal of providing direct-access while enforcing data
protection, failure atomicity, consistency, as well as the scalability to multi-cores.
\subsection{Overview of KucoFS}
\label{overview}
\begin{figure}
	\centering
	\includegraphics[width=1\linewidth]{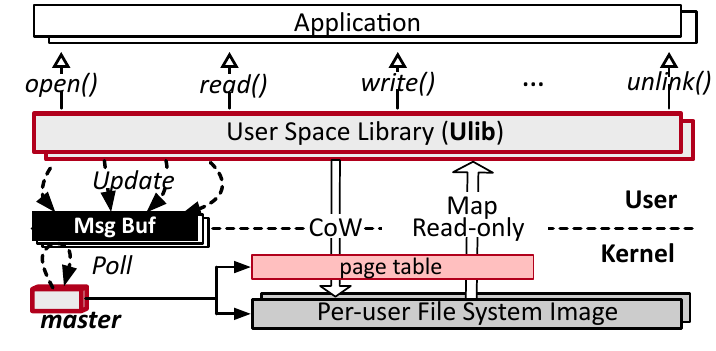}
	\vspace{-0.35in}
	\caption{Architecture of KucoFS.}
	\label{kucofs}
	\vspace{-0.25in}
\end{figure}

% Overall Architecture
Figure~\ref{kucofs} shows the architecture of KucoFS.
KucoFS consists of an user-level
library and a global kernel thread, which are
respectively called \emph{Ulib} and \emph{master}. \emph{Ulib} communicates with
the \emph{master} via a exclusively owned message buffer.
% , \emph{Ulib} abstracts a group
% of POSIX compatible interface (e.g., \texttt{open/read/write/unlink},
% etc), so the applications can access the file system without any modification
% to the source code.
In KucoFS, each user owns a partition of file system image, which is mapped into
user-level address space with read-only access rights.
By linking with \emph{Ulib}, an application can post memory \texttt{Load} instructions
to directly locate the data for those read-only operations (e.g., \texttt{read/stat}).
\emph{Ulib} writes file data by always indirecting updates to new data
pages with a copy-on-write mechanism. To enable user-level direct write,
the \emph{master} modifies the permission bits in the page table to switch
the newly allocated data pages between ``writable'' and ``read-only'' when \emph{Ulib} is updating them.
\emph{Ulib} is not allowed to update metadata directly.
Instead, it posts a request to the \emph{master} through the message buffer,
and the \emph{master} updates the metadata on behalf of it.
% Similar to those kernel file systems, a file descriptor table (describing the status
% of each opened file) is maintained in the \emph{Ulib}.
% Note that the file data are out-of-place updated in
% userspace directly but the corresponding metadata are modified by
% the kernel thread.

% Data Layout
KucoFS adopts both DRAM and NVM to manage the file system image
(see Figure~\ref{fig:layout}).
For efficiency, KucoFS only operates on the DRAM data for normal
requests. In DRAM, an array of pointers (\emph{inode table})
is placed at a predefined location to point to the actual \emph{inode}s.
The first element in the \emph{inode table} always points to the \emph{root inode} of each user,
therefore, \emph{Ulib} can lookup iteratively from the root inode to any file directly in user space.
KucoFS uses an Ext2-like~\cite{ext2} \emph{block mapping} to map a file to its
data pages. We choose block mapping, instead of the widely used extent tree, to support lock-free fast read (in Section~\ref{subsec:read}).
% It consists of both direct blocks and indirect blocks (storing
% pointers to direct blocks), so as to support large files.
We then introduce skip-list~\cite{skiplist} to organize the \emph{dentry list}
of each directory, so as to achieve atomicity and consistency (in Section~\ref{subsec:offloading}).

To ensure the durability and crash consistency of metadata, KucoFS further
places an append-only persistent operation log in NVM.
When the \emph{master} updates the metadata, it first atomically appends a log entry,
and then actually updates the in-memory metadata.
To avoid the operation log from growing arbitrarily, the \emph{master} will
periodically checkpoint the modifications to the NVM metadata pages in the background
and finally truncate the log (in Section~\ref{subsec:recovery}).
Since the operation log only contains light-weight metadata, such checkpoint overhead is not high.
In face of system failures, the in-memory metadata can always be recovered
by replaying the log entries in the operation log.

% Besides, by atomically updating the skiplist at proper time point,
% the user space applications are insured to always have a consistent view of
% the file system image.
In addition to the operation log and metadata pages, the extra NVM space is cut into
contiguous 4~KB \emph{data pages} to store the file data.
%Worth noticing, the in-memory metadata is updated
%without any synchronization overhead (e.g., locking or atomic operation) because all
%the metadata updates are delegated to a single kernel thread.
The free data pages are managed with both a bitmap in NVM and a
free list in the DRAM (for fast allocation). Similar to the metadata pages,
the bitmap is also lazily persisted by the \emph{master} during the checkpoint.
% Once a system crash occurs, it still can be recovered from the remaining operation log.
% How to manage the file descriptor?

\subsection{Metadata Management}
\label{subsec:offloading}

\begin{figure}
	\centering
	\includegraphics[width=1\linewidth]{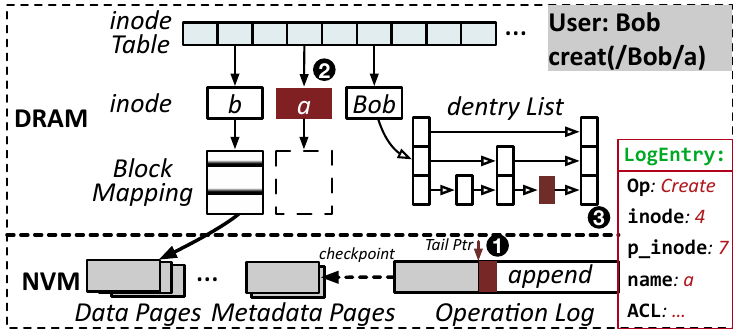}
	\vspace{-0.3in}
	\caption{Data layout in KucoFS and the steps~to~create~a~file.}
	\label{fig:layout}
	\vspace{-0.2in}
\end{figure}

% To prevent a malicious program from corrupting the file system,
% all the metadata updates are delegated to the \emph{master}.
% However, following such centralized architecture, KucoFS is hard
% to scale on a multi-core platform since a single kernel thread
% will be the bottleneck.
%With the centralized design in KucoFS, the processing power of the \emph{master}
%determines the maximum throughput that KucoFS can achieve.
KucoFS delegates all metadata updates to the \emph{master}.
To relieve the pressure of the \emph{master}, we propose to
1)  minimize its metadata indexing overhead with \emph{index offloading}
and 2) reduce the metadata persistence overhead with batching.
% The centralized design of \emph{master} will cause scalability issue since
% the maximum throughput that KucoFS can achieve will be limited by the
% processing power of \emph{master}.

\noindent
\textbf{Index Offloading.}
To update metadata, the \emph{master} needs to perform iterative pathname resolution
from \emph{root inode} down to the directory containing this file.
%Such high indexing overhead largely restricts the efficiency of \emph{master}.
When a large number of processes access concurrently, such indexing
overhead is a heavy burden for the master.
Things become even worse when a directory contains
a large number of sub-files or the file path is long.
To address this issue, we propose to offload the pathname resolution from the
\emph{master} to \emph{Ulib}.

By mapping the file system image to user space,
\emph{Ulib} is enabled to locate the related metadata directly in user-level before posting a
metadata update request. Take \texttt{creat}
operation for example, \emph{Ulib} finds the address of the predecessor in
its parent directory's \emph{dentry list}. It then posts the request
to the \emph{master} by piggybacking the addresses of the related metadata. In this way,
the \emph{master} can directly insert
a new \emph{dentry} into the \emph{dentry list} with the giving address. The addresses of both the \emph{dentry}
in the parent directory and \emph{inode} itself are provided for the \texttt{unlink}
operation.
However, we still need extra techniques to ensure the correctness:

% Insure Correctness
First, we need to ensure that \emph{Ulib} can always read consistent directory tree
when the \emph{master} is updating them concurrently.
To address this issue, we organize the \emph{dentry list} of each directory with a
skip-list~\cite{skiplist} and the key is the hash value of each file name.
Skip-list is a linked list-like data structure with multiple layers, and each higher layer
acts as an ``express lane'' for the lists below, thus providing $O(logN)$ search/insert complexity (see Figure~\ref{fig:layout}).
More importantly, by performing simple pointer manipulations on a singly linked list with
CPU's atomic operations, we can atomically update the list.
We enforce the \emph{master} to updates the \emph{dentry list} at
different time point for different operations: For \texttt{creat}, it
inserts a new \emph{dentry} on the final step, to atomically make the created file visible;
For \texttt{unlink}, it deletes the \emph{dentry} firstly.
Hence, \emph{Ulib} is guaranteed to always have a consistent view of the directory tree even without
acquiring the lock.
% such properties greatly simplify the way to achieve fast and consistent ``direct access''
% in KucoFS (described later).
Renaming involves updating two \emph{dentries} simultaneously, so it is possible for
a program to see two same files at some time point.
To address this issue, we add an \emph{dirty} flag in each \emph{dentry} to 
prevent the \emph{Ulib} from reading such inconsistent state.

Second, it's possible that the pre-located metadata by \emph{Ulib} becomes
obsolete before it is actually accessed by the \emph{master} (e.g., the \emph{inode}
or \emph{dentry} has already been deleted by the \emph{master} for other concurrent processes).
%To guarantee that the \emph{master} can correctly update the metadata,
%we further introduce the following techniques:
To solve this problem, we reuse the \emph{dirty} bit in each \emph{inode}/\emph{dentry}.
Once an item is deleted, this bit is set to an invalid state.
Therefore, other applications and the \emph{master} itself
can determine the liveness of each metadata.
The deleted items are temporarily kept in place and reclaimed via
an epoch-based reclamation mechanism (EBR)~\cite{fraser2004practical}.
We follow a classic way by using three reclamation queues, each of which
is associated with an epoch number. The \emph{master} pushes the deleted items
only to the current active epoch queue. When all \emph{Ulib} instances are active in the current epoch,
the \emph{master} then increases the global epoch and begins to reclaim the space from the oldest queue.
A reader executing in user level can suffer arbitrary delays due to thread scheduling,
impacting the reclaim efficiency. However, we believe it is not
a serious issue since KucoFS only reclaims these obsolete items periodically.

Third, a pre-located \emph{dentry} may no longer be the predecessor
when a new \emph{dentry} is inserted between them. Hence, the \emph{master} also
needs to check the legality of the pre-located metadata by comparing the related fields.
Note that the \emph{master} can update in-memory metadata without any synchronization overhead (i.e.,
locking) since all the metadata updates are delegated to the \emph{master}~\cite{sosp17ffwd}.
% However, it is possible that the pre-located metadata has been already deleted
% before the kernel thread accessing them. To insure that the kernel thread
% correctly
% we
% Further offloading.

\ul{\emph{Examples.}} To create a file, \emph{Ulib} sends a \texttt{creat} request to the \emph{master}
to create a file. The address of the predecessor in its parent directory's
\emph{dentry list} is put in the message too. Upon receiving the request, the \emph{master} does
the following steps (as shown in Figure~\ref{fig:layout}):
\ding{202} reserves an empty inode number from the
\emph{inode table} and appends a log entry to guarantee crash consistency.
This log entry records the inode number, file name, parent directory inode number, and other
attributes; \ding{203} allocates an \emph{inode} with each field filled,
and updates the \emph{inode table} to point to this \emph{inode}, and  \ding{204}
inserts a \emph{dentry} into the \emph{dentry list} with the given address, to make the created file visible.
To delete a file, the \emph{master} appends a \emph{log entry} firstly, deletes the
\emph{dentry} in the parent directory with the given addresses, and finally frees the related spaces
(e.g., \emph{inode}, NVM file pages and \emph{block mapping}). With such strict execution order, the
failure atomicity and consistency (described in Section~\ref{sec:goal}) is guaranteed.

\noindent
\textbf{Batching-based Metadata Logging.}
The \emph{master} ensures the crash consistency of metadata by appending log entries
and flushing them out of the CPU cache. However, cache flushing
leads to significant overhead since NVM has poor write bandwidth.
Fortunately, the \emph{master} serves many user space
applications and it can flush log entries with batching.
Following this idea, we let the \emph{master} fetch multiple requests at a time
from concurrent applications and process them in a batch manner.
multiple log entries from different requests now can be merged into a large log entry.
After it is persisted, the \emph{master} then updates the in-memory metadata one-by-one with
the order described above, and finally sends the acknowledgments back.
Such processing mode has the following advantage:
CPU flushes data with cacheline granularity (typically 64~B), which is larger than most of the
log entries, by merging and persisting them together, the number of flushing operations is dramatically
reduced.
Note that the aforementioned batching is different from Aerie and Strata:
Aerie batches requests before sending them to the TFS,
so as to reduce the cost of posting RPCs.
The KernFS in Strata digests batches of operations from the log, which coalesces
adjacent writes and forms sequential writes.
Instead, KucoFS batches log entries to amortize the data persistence overhead,
leveraging the mismatch between the flush granularity and the log entry size.

\subsection{Write Protocol}
\label{subsec:write}
Another key design principle lies in \emph{how to provide
efficient, consistent and safe write protocol}. To achieve these goals,
we propose \emph{pre-allocation} and \emph{direct-access range-lock} to
simplify the way of interaction between \emph{Ulib} and \emph{master}.

Similar to NOVA~\cite{fast16nova} and PMFS~\cite{eurosys14aerie},
we use a copy-on-write (CoW) mechanism to
update data pages. It updates file pages by moving
the unmodified part of data from the old place as well as the application data to new data pages.
CoW causes extra copying overhead for small-sized updates. In most cases, however,
it dismisses the double write overhead as in
redo/undo logging and the log cleaning overhead as in log-structured
data management.

KucoFS first uses CoW to update the data pages, and then atomically appends
a log entry to record the metdata modifications, during which the old data and metadata is never touched.
Once a system failure occurs before a write operation is finished, KucoFS simply rollbacks
to its original state. As such, the failure atomicity and consistency is guaranteed.
We rely on the \emph{master}
to enforce write protection over each file page leveraging the permission bits in the page table:
When the user-level programs directly write data, the \emph{master} carefully manipulates the
permission bits of the related data pages.
% To enable direct but safe access in user space, the \emph{master} is devoted to switches
% the permission bits of the related data pages between ``writable'' and ``read-only''.
% It's equally important to reduce the overhead of \emph{master} in file writing.
An intuitive write protocol is:
\begin{compactenum}[1)]
\item \emph{Ulib} sends the first request to the \emph{master} to lock the file, reserve free data pages
and make them ``writable'';
\item \emph{Ulib} relies on CoW to copy both the unmodified data from the old place and new data from the
user buffer to the reserved data pages, and flush them out of the CPU cache;
\item \emph{Ulib} sends a second request to the \emph{master} to reset the newly written
data pages to ``read-only'', append a new log entry (inode number, offset,
size and related NVM addresses) to the operation log, update the metadata (i.e.,
\emph{inode}, \emph{block mapping}) and finally release the lock.
\end{compactenum}

\begin{figure}
  \centering
  \includegraphics[width=1\linewidth]{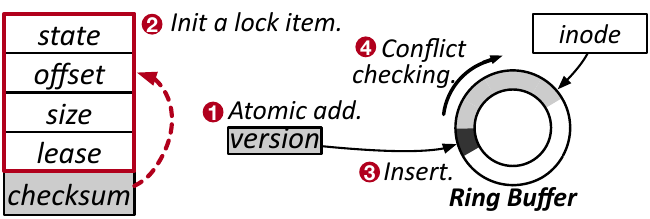}
  \vspace{-0.3in}
  \caption{Layout of Direct Access Range-Lock.}
  \label{fig:lock}
  \vspace{-0.25in}
\end{figure}

% \noindent
% $\blacktriangleright$ \emph{Consistency}.
% However, for a unaligned written page, one needs
% to copy the old page to a new place and write the unaligned data to form a new
% page. Such redundant copying overhead dramatically impacts the write efficiency
% especially for small-sized I/Os~\cite{sosp17strata}. In KucoFS, we propose to
% address this issue with a per-page log, which will be described latter.

% \noindent
% $\blacktriangleright$ \emph{Fine-grained Concurrency Control}. Most kernel file systems
% synchronizes the \emph{syscall}s of file I/O with coarse-grained lock,
% so concurrent file accesses to the same file are strictly serialized.
% For applications such as in high-performance computing~\cite{}, however,
% supporting more fine-grained concurrency control is extremely important, and
% this demand will continue to grow on emerging rack-scale computers that have
% hundreds to thousands of processors~\cite{}.

%\noindent
%$\blacktriangleright$ \emph{Security}. By offloading the memory copy overhead,
%we are forced to allow \emph{Ulib} to directly write the NVM space. We believe
%this is still tolerable since \emph{Ulib} only writes file data to the newly
%allocated space with CoW mechanism instead of modifying the existing file system
%image directly. Besides, the modifications are made visible only after the
%metadata is updated by \emph{master}.
%
%This doesn't violate with our security
%requirement since the newly appended log entries is validated only after
%the \emph{master} modifies the persistent tail pointer.

We can observe that a single \texttt{write} operation involves posting two
requests to the \emph{master}. This can not only lead to high write latency,
but also limit the efficiency of the \emph{master} since it is frequently
involved. Thus, we propose \emph{pre-allocation} and \emph{direct access range-lock}
to avoid sending the first request to the \emph{master}.

\noindent
\label{subsubsec:early-write}
\textbf{Pre-allocation.}
Rather than posting the request to the \emph{master}
to reserve free pages for each \texttt{write} operation, we allow \emph{Ulib} to
lazily allocate data pages from the \emph{master} (4~MB at a time in our implementation).
These data pages are managed privately by \emph{Ulib} with a free list.
When an application exits, the unused data pages are given back to the \emph{master}.
For an abnormal exit, these data pages are temporarily non-reusable by other applications,
but still can be reclaimed after rebooted by replaying the operation log.
% We also allow \emph{Ulib} to reserve space in
% per-page log directly by atomically modifying a mirrored volatile tail pointer.

%Besides,
%We can notice that \emph{Ulib} only needs to post one
%request to update the metadata, so the latency is reduced. By moving the data copy
%out of the critical path, the concurrency is improved too.

\noindent
\textbf{Direct Access Range-Lock.}
To completely avoid sending the first request as described in the naive write protocol,
we further propose \emph{direct access range-lock}. It coordinates the concurrent writes
directly in user-level, since we cannot rely on a \emph{master} to acquire the lock anymore.

%``Out-of-place'' writing file data without any coordination still cause inconsistency issue.
%Considering two applications (e.g., $\rm P_1$ and $\rm P_2$) that
%concurrently update the same file with data range of $\rm [a,d]$ and $\rm [b,c]$
%(where $\rm a<b<c<d$): As shown in Figure~\ref{inconsistency}, both $\rm P_1$
%and $\rm P_2$ update four data pages (the middle two pages are completed
%overwritten while the other two is partially updated). Following a CoW approach,
%it's possible that the latter application (e.g., $\rm P_2$) fails to see the written data by
%$\rm P_1$ and still copies the old data to form the
%new pages, thus leading to the incorrect state as in case (c).
% \emph{Ulib} needs to reserve spaces in the per-page log of both the first and the
% forth page, so the atomicity of space reservation is not guaranteed
% and may cause data overlapping (case (c)). Besides, the log reservation
% order between $\rm P_1$ and $\rm P_2$ may be inconsistent with the order that \emph{master}
% updates the metadata (see case (d)).

As shown in Figure~\ref{fig:lock}, we assign each opened file a range lock (i.e., a DRAM ring buffer),
which is pointed by the \emph{inode}.
 \emph{Ulib} writes a file
by acquiring the range-lock first, and the file writing is delayed once a lock conflict occurs.
Each slot in the ring buffer has five fields, which are \emph{state,
offset, size, lease} and a \emph{checksum}.
The \emph{checksum} is the hash value of the first four fields.
We also place a \emph{version} at the head of
each ring buffer to describe the ordering of each write operation.
To acquire the lock of a file, \emph{Ulib}
firstly increments its \emph{version} with atomic \texttt{fetch\_and\_add}. It then
inserts a lock item into a specific slot in the ring buffer, and the location
is determined by the fetched \emph{version} (modulo the ring buffer size).
After this, \emph{Ulib} traverses
the ring buffer backward to find the first conflicting lock item (i.e., their written
data overlaps). If it exists, \emph{Ulib} verifies its \emph{checksum},
and then polls on its \emph{state} until it is released.
\emph{Ulib} also checks its \emph{lease} field repeatedly to avoid deadlock if an
application is aborted before it releases the lock.
Once the lock has been required, \emph{Ulib} process the second and third steps
 described in the naive protocol. In step 3, the \emph{version} is encapsulated in the
 requests, so the \emph{master} can persist it in the log entry.

% to describe this write operation (e.g.,
%\emph{inode} number, offset, size and NVM addresses).

%We allow the applications to asynchronously wait for the response message and
%continue executing the following code.
%When a subsequent request that conflicts with this \texttt{write} (access to
%the same file), \emph{Ulib} blocks the application until the corresponding response message is returned.
%Therefore,
%by posting a subsequent \texttt{fsync}, the semantic of traditional paradigm
%\emph{``buffered write followed by a fsync''} still holds in KucoFS.
Worth noticing, our proposed range-lock supports concurrent writing in the same file covering different data pages.
Such fine-grained concurrency control is important
in high-performance computing~\cite{tocc16yildirim,
blanas2014parallel} and the emerging rack-scale computers with hundreds to thousands of cores~\cite{keeton2015machine}.

% One limitation is that copy-on-write mechanism can cause redundant copying
% overhead for small-sized I/Os (NOVA also encounters the same problem). We
% believe it will be a feasible approach to introduce an append-only log
% alongside each data page to accept the unaligned updates.
% KucoFS has not implemented memory-mapped I/O yet. We leave them as
% our future work. \cym{TLB?}
% Besides,
% the watchdog in Linux kernel prevents a kernel thread from
% being long-term executed by sending.

\noindent
\textbf{Write Protection.}
%KucoFS only allows the \emph{master} to update the metadata, so
%All the metadata in KucoFS is only writable by the \emph{master}
%By offloading the memory copy overhead to userspace, we are forced to allow \emph{Ulib} to
KucoFS strictly controls the access rights to the file system image:
Both in-memory metadata and the persistent operation log are critical
to the file system, so the \emph{master} is the only one that is allowed to update them.
\emph{Ulib} only has write access to its privately managed free data pages. However,
these pages are immediately changed to ``read-only'' once they are allocated
to serve the \texttt{write} operations.
Since both the metadata and valid data pages are non-writable,
KucoFS is immune to arbitrary memory writes.
However, there are still two anomalies:
1) the private data pages still can be corrupted within a write operation
by other concurrent threads. However, a kernel-based file system cannot
cope with such case either~\cite{eurosys14pmfs}, and we believe this is unlikely to happen.
2) a buggy application can still corrupt the range lock or the message buffer,
since they are directly writable in user space.
We add \emph{checksum} and \emph{lease} fields at each slot, enabling the user-level programs
to identify whether the inserted element has been corrupted.
Both the lock item and request message only contains a few tens of bytes of data,
so the hash calculating overhead is not high. Besides,
The secret key for generating the checksum is owned by the \emph{master} and
granted only to those trusted applications.
Therefore, KucoFS is even immune to some malicious attacks (e.g., replay or DoS),
though this is not the main target of this paper.

When the \emph{master} updates the page table for each write operation,
it needs to explicitly flush the related TLB entries
to make the modifications visible. This indicates that each \texttt{write}
operation in KucoFS involves twice of TLB flushing. Luckily,
we can allocate multiple data pages at a time in \emph{pre-allocation} phase,
So the TLB entries can be flushed in batch, which reduces the flushing overhead dramatically.

\subsection{Read Protocol}
\label{subsec:read}

%Following a CoW mechanism as in the write protocol, the existing pages will not be modified since
%\emph{Ulib} always redirects the data to new places.
KucoFS updates data pages with CoW mechanism, hence, any data page is in either old or new version.
This provides us the opportunity to design an efficient read protocol to directly read in user-level.
Considering that the \emph{master} may be updating the metadata for other concurrent writers,
the main challenge is how to
read a consistent \textbf{snapshot} of \emph{block mapping}s efficiently despite other concurrent writers.
%One approach is to change
%the range-lock in Section~\ref{subsubsec:early-write} to a read-write lock. However,
%the lock service is too heavy-weight for \texttt{read} operation since it involves
%expensive atomic operation and inter-core communication.

\begin{figure}
	\centering
	\includegraphics[width=1\linewidth]{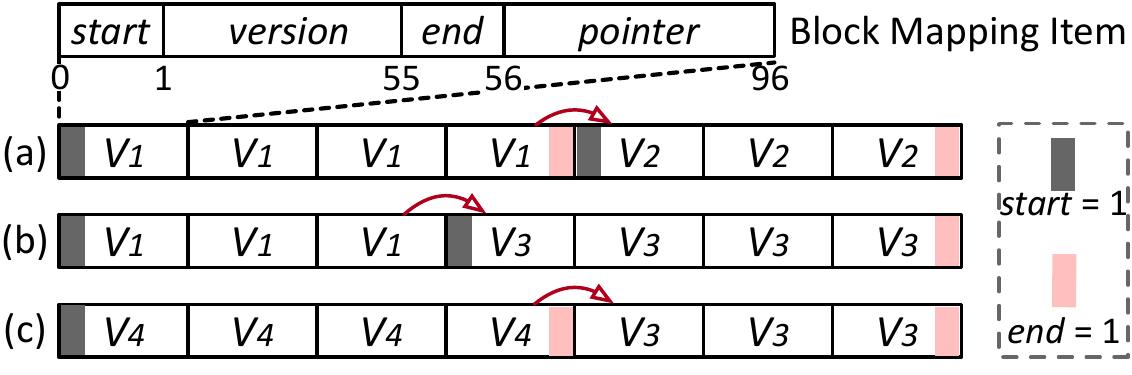}
	\vspace{-0.35in}
	\caption{Lock-Free Fast Read with Version Checking.}
	\label{fast-read}
	\vspace{-0.25in}
\end{figure}

Hence, we propose
\emph{lock-free fast read}, which guarantees that readers never read data from unfinished writes.
It achieves this
by embedding a \emph{version} field in each pointer of the \emph{block mapping}:
As shown in Figure~\ref{fast-read}, each 96-bit block mapping item contains four
fields, which are \emph{start}, \emph{version}, \emph{end}, and \emph{pointer}.
% Take a \texttt{write} operation with three updated data pages for example,
% when all the preliminary work is completed, \emph{Ulib} sends a remote call to
% the \emph{master} and the \emph{version} (in Section~\ref{subsubsec:early-write})
% is piggybacked in the request as well.
Take a \texttt{write} operation with three updated data pages for example,
when the \emph{master} updates the \emph{block mapping},
the header of three mapping items are constructed with the following layout:
\setlength{\fboxsep}{0cm}
\fbox{\colorbox{gray}{$\rm 1_{}$}$|$\colorbox{white}{$\rm V_1$}$|$\colorbox{white}{$\rm 0_{}$}}
\fbox{\colorbox{white}{$\rm 0_{}$}$|$\colorbox{white}{$\rm V_1$}$|$\colorbox{white}{$\rm 0_{}$}}
\fbox{\colorbox{white}{$\rm 0_{}$}$|$\colorbox{white}{$\rm V_1$}$|$\colorbox{pink}{$\rm 1_{}$}}.
Note that all the three items share the same \emph{version} (i.e., $\rm V_1$), which
is provided by \emph{Ulib} when it acquires the range lock (in Section~\ref{subsec:write}).
The \emph{start} bit of the first item and the \emph{end} bit of the last item are set to 1.
We only reserve 40-bit for \emph{pointer} field since it always points to a
4~KB-aligned page (the lower 12 bits can be discarded). It's easy to understand that
when there are no concurrent writers, the block mapping items should satisfy any of
the conditions in Figure~\ref{fast-read}:
\begin{compactenum}[(a)]
\item \emph{Writes without overlapping.} The items with the same \emph{version} are enclosed with a \emph{start} bit and
an \emph{end} bit, indicating that multiple threads have updated the same file but different data pages.
\item \emph{Overlapping in the tail.} The reader sees a \emph{start} bit when the \emph{version} increases, indicating that
a thread has overwritten the end part of the pages that are updated by a former thread.
\item \emph{Overlapping in the head.} The reader sees an \emph{end} bit before the \emph{version} decreases, indicating that
a thread has overwritten the front part of the pages that are updated by a former thread.
\end{compactenum}

If \emph{Ulib} meets any cases that violate the above conditions, we assert that the
\emph{master} is updating the block mappings for other concurrent write
threads. In this case, \emph{Ulib} needs to reload the metadata again and checks its validity.
To reduce the overhead of retrying, the read thread copies the file data to user's buffer
only after it has successfully collected a consistent version of the block mapping. This
is achievable because the obsolete NVM pages are lazily reclaimed.
When the modified mapping items span to multiple cachelines,
the \emph{master} also adds extra \emph{mfence} to serialize the updates.
By this way, the read threads can see the updates in order.

%We believe a fine-grained read protection mechanism can be achieved by partitioning the
%NVM space into different regions and selectively mapping them to user level with different read access rights.

\noindent
\textbf{Read Protection.} Leveraging the permission bits to
enforce read protection is more challenging, since metadata have semantically
richer permissions~\cite{eurosys14aerie}. Hence,
instead of maintaining a fully-compatible hierarchical/group access control as in kernel-based file systems,
we partition the directory tree into per-user sub-trees and each user has a private root node.
When a program access KucoFS, only the sub-tree (i.e., \emph{inode table}, \emph{inodes},
\emph{dentry lists}, etc.) and the related data pages of the current user are mapped to
its address space, while other space is invisible to it. To alleviate the bookkeeping overhead
for page mapping, the \emph{master} assign each user 4~MB of contiguous DRAM/NVM blocks, which forms
the per-user file system image (i.e., DRAM metadata, operation log, data pages, etc.).
Similar to Arrakis~\cite{osdi14arrakis},
KucoFS doen't provide a explicit way for data sharing between different users,
yet there are several practical approaches:
1) create a standalone partition that every users have read/write access to it;
2) issue user-level RPCs to a specific user to acquire the data.
We believe such tradeoff is not likely to be an obstacle to its application in real-world scenarios,
since KucoFS naturally supports efficient sharing between applications within the same user,
which are the more common case.

%\noindent
%\textbf{Discussion.}
%Similar to Aerie, we enforce read protection and ACLs directly in \emph{Ulib} with a software approach,
%since this is not the main task in this paper.
%We still have two alternative ways to implement read protection via a hardware approach:
%(1) Grouping the data pages that belong to different users and assign them with different
%read permissions in the page table.
%(2) Intel deployed a new hardware feature named \emph{memory protection keys} (MPK)~\cite{mpk}.
%It enforces thread-local permission control to each group of memory pages,
%without requiring modification of page tables.
%Through the hardware approach, a non-root program can no
%longer read the data from a root user. PMK is only supported
%by Intel Xeon Scalable family and we have no access to such
%CPU, so we leave this part as our future work.
%\cym{Per-user protection.}

% which exhibits asymmetric performance for inter(intra)-processor communication,
% local/remote memory access
\subsection{Log Cleaning and Recovery}

We introduce a checkpoint mechanism to avoid the operation log from
growing arbitrarily: When a \emph{master} is not busy or the size of operation
log grows out of a maximum size, it periodically
applies the metadata modifications to NVM metadata pages by replaying log entries in the
operation log. The bitmap that used to manage the NVM free space is updated and persisted
as well. After that, the operation log is truncated.
Each time KucoFS is restarted,
the \emph{master} first replays the un-checkpointed log entries in the operation
log, so as to make the NVM metadata pages up-to-date.
It then copies the NVM metadata pages to DRAM. The free list of NVM data
pages is also reconstructed according to the bitmap stored in NVM.
Keeping redundant copies of metadata between DRAM and NVM can introduce higher consumption
of NVM/DRAM space. But we believe it is worth the efforts because by selectively placing the
(un)structured metadata in DRAM and NVM, we can perform fast indexing directly in DRAM,
append log entries with reduced persistency overhead (batching), and lazily checkpoint
in the background without affecting performance. As our future work,
we plan to reduce the DRAM footprint by only keeping the metadata of active files
in DRAM.

% As mentioned in Section~\ref{overview}, KucoFS also maintains a copy of metadata in NVM
% space, which has the same layout as that in DRAM. The \emph{master} will periodically
% apply the metadata modifications to NVM space by replaying the operation log.
% The bitmap that used to manage the NVM free space will be persisted
% as well. After that, the operation log is truncated.
% \cym{meta cache size.}
%
%For those applications with less consideration on data security, we also provide aggressive but
%optional optimizations to further accelerate the \texttt{write} operation.
%\emph{Ulib} is allowed to directly update the in-memory metadata, so \emph{master} only need to
%append a log entry to the operation log for crash consistency

%(1) for overwritten pages, \emph{Ulib} only writes file data to the newly allocated space;
%(2) for partially updated pages, \emph{Ulib} appends the written data at the end of a per-page log.

\subsection{Examples: Putting it all together}
\label{subsec:recovery}
We finally summarize the design of KucoFS by walking through an example of writing 4~KB of data
to a new file and then reading it out. First of all, this program links with \emph{Ulib} to map the
related NVM/DRAM space of the current user into its address space.

\noindent
\textbf{Open.}
% An application is required to link with \emph{Ulib} and call its API to access KucoFS (in Section~\ref{sec:impl}).
Before sending the \texttt{open} system call, \emph{Ulib} pre-locates the related metadata first. Since this is
a new file, \emph{Ulib} cannot find its inode. Instead, it finds the predecessor in its parent
directory's dentry list for latter creation.
The address, as well as other information (e.g., file name, \texttt{O\_CREAT} flags, etc.)
are encapsulated in the \texttt{open} request.
When the \emph{master} receives the request, it creates this file based on the given address. It also allocates a
range-lock ring buffer for this file since it's the first time to open it. Then, the \emph{master} sends
a response message. After this, \emph{Ulib} creates a file descriptor for this opened file and
returns back to the application.

\noindent
\textbf{Write.}
The application then uses \texttt{write} call via \emph{Ulib} to write 4~KB of data to this created file.
First, \emph{Ulib} finds the inode of this file and locks it with the direct access range-lock.
\emph{Ulib} blocks the program when there are write conflicts and wait until the corresponding lock has been released. After this,
\emph{Ulib} can acquire the lock successfully. It then allocates a 4~KB-page from its privately managed
lists, copies the data into it, and flushes them out of CPU cache.
\emph{Ulib} needs to post extra request to the \emph{master} to
allocate more free data pages once its own space is used up. Finally, \emph{Ulib}
sends the \texttt{write} request to the \emph{master} to perform the loose ends,
which includes: change the permission bits of the written data pages to ``read-only'', atomically
appending a log entry to describe this write operation, update the in-memory metadata,
and finally unlock the file.

\noindent
\textbf{Read.}
KucoFS enables reading file data without interacting with the \emph{master}. To read the first 4~KB from this file,
\emph{Ulib} directly locates the inode in user space and reads the first block mapping item (i.e., the pointer).
The version checking is performed to ensure its state satisfies one of the three conditions described
in Section~\ref{subsec:read}. After this, \emph{Ulib} can safely read the file data page pointed by the \emph{pointer}.

\noindent
\textbf{Close.} \emph{Ulib} also needs to send a \texttt{close} system call to the \emph{master} upon closing this file.
The \emph{master} then reclaims the space of the range lock ring buffer if no other processes is accessing this file.

% + Naive Mode: Lock and Reserve - Write - Update Meta

% + Lazily Reserve Space. Write - Update Meta and Write Log.

% + Page Logging

% + Parallel Write

% + Range Lock

% * For file write, we interleave the Syscall sending and file data copying (from
% user buffer to FS image).
% * More fined-grained locking: allow concurrent access to the same file. We also
% carefully moves most of the heavy works out of the lock to increase the concurrency.
% * Kernel thread persists log entries with batching

% \begin{algorithm}
%     \SetKwInOut{Input}{Input}
%     \SetKwInOut{Output}{Output}
%
%     %\underline{function pwrite} $(fd, offset, size)$\;
%     \Input{fd, offset, size}
%     $pages = (size < 4KB) ?  1 : size / 4KB$\;
%     $head = tail = 0$\;
%     \If {$offset \% 4KB != 0$}
%     {
%        pages -= 1;
%        head = 1\;
%     }
%     \If {$(offset + size) \% 4KB != 0 \& size >= 4KB$}
%     {
%        pages -= 1;
%        tail = 1\;
%     }
%     %\Output{$\gcd(a,b)$}
%
%     \If{$P_a!=null$}
%     {
%        \tcc{\small{Alloc free pages for overwrite.}}
%        pages = alloc(sizeof($P_a$) * PAGE\_SIZE)\;
%     }
%     \If{$P_{head}!=null$}{
%        \tcc{Reserve space in log}
%
%     }
%     \caption{Euclid's algorithm for finding the greatest common divisor of two
%     nonnegative integers}
% \end{algorithm}

%\subsubsection{}

% \subsubsection{Per-Page Logging}

% * Optimizations for partial update (extra coping overhead brought by the CoW
% mechanism) by introduce a per-page log.

% \subsubsection{Range Lock}

\section{Implementation}
\label{sec:impl}

% KucoFS is implemented without any modifications to existing Linux kernel.
KucoFS is implemented into two parts: a loadable kernel module (i.e., the \emph{master})
and a shared library (i.e., the \emph{Ulib}). Each \emph{Ulib} instance communicates with the
\emph{master} with an exclusively owned message buffer.
% Once the kernel module is
% inserted into kernel, it will launch the \emph{master} threads and perform the
% initialization works.

%
%\noindent
%\textbf{Kernel/Userspace Communication.}
%%borrow the idea from FlexSC, but has difference.
%To support fast and concurrent communication between \emph{Ulib} and
%\emph{master}, each \emph{Ulib} instance
%maintains an exclusively owned message buffer, which consists of multiple request
%slots. \emph{Ulib} sends a request by copying it into one of the message
%slots in a round-robin fashion. Each message contains a request type (e.g.,
%\texttt{creat/write}, etc.), the corresponding parameters and a return value. The
%\emph{master} scans all message buffers of the active \emph{Ulib}s to find new
%requests. Once a new request is processed, the \emph{master} writes the
%return value to the corresponding message slot. For a synchronous request
%(e.g., metadata update), \emph{Ulib} polls until it gets the return value.
\begin{figure*}
	\centering
	\includegraphics[width=1\linewidth]{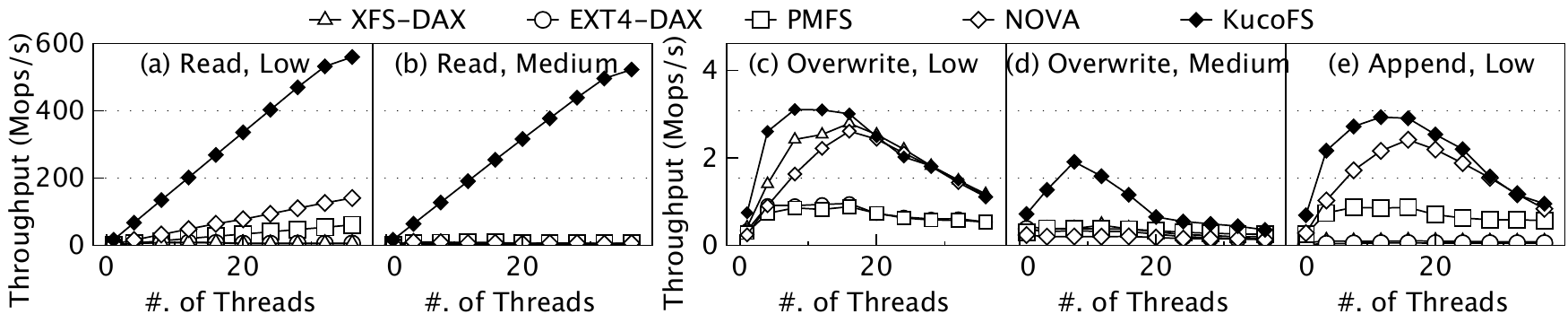}
	\vspace{-0.25in}
	\caption{Read and write throughput with FxMark. (\emph{``Low'': different threads read(write)
			data from(to) different files; ``Medium'': in the same file but different data blocks;
			We use default I/O size of 4~KB.})}
	\label{fxmark-rw}
	\vspace{-0.15in}
\end{figure*}

\noindent
\textbf{KucoFS's APIs.}
KucoFS provides a POSIX-like interface, so existing applications are enabled to
access it without any modifications to the source code. It achieves this by setting the \texttt{LD\_PRELOAD} environment variable. \emph{Ulib} intercepts all APIs in
standard C library that are related to file system operations.
\emph{Ulib} processes the \emph{syscall} directly if the prefix of the accessed
file matches with a predefined string (e.g., \emph{``/kuco''}). Otherwise, the
\emph{syscall} is processed in legacy mode. Note that \texttt{write} operations
only pass the file descriptors to locate the file data, therefore, \emph{Ulib}
distinguish the \texttt{write} operations from legacy file systems by only using
big file descriptor numbers (greater than $2^{20}$ in our implementation).
% Besides,
% \emph{Ulib} should be thread safe since some resources need to be carefully shared
% among all the threads in the same process: For concurrent \texttt{open} operations, we
% reduce competition on unused file descriptor numbers by partitioning them to
% different threads. We further partition the message buffer so that different
% threads in the same process can write the requests to their own request slots.

\noindent
\textbf{Memory-mapped I/O.}
Supporting DAX feature in a copy-on-write file system needs extra
efforts, since the files are out-of-place updated in normal \texttt{write} operations~\cite{fast16nova}.
Besides, DAX leaves great challenges for programmers to correctly use NVM space with atomicity and
crash consistency. Taking these factors into consideration, we
borrow the idea from NOVA to provide \texttt{atomic-mmap}, which has higher consistency guarantee.
When an application maps a file into user space, \emph{Ulib} copies the file data to its
privately managed data pages, and then sends a request to the master to map these pages into
contiguous address space. When the application issues a \emph{msync} system call, \emph{Ulib} then handles it as a
write operation, so as to atomically makes the updates in these data pages visible to other applications.

\section{Evaluation}
\label{sec:eval}
In this section, we evaluate the overall performance of KucoFS with micro(macro)-benchmarks and real-world applications. We also learn the effects brought by its internal mechanisms.

\subsection{Experimental Setup}
\label{subsec:testbed}
\textbf{Testbed.}
Our experimental testbed is equipped with 2$\times$ Intel Xeon
Gold 6240M CPUs (36 physical cores and 72 logical threads), 192 GB DDR4 DRAM,
and six Optane DC persistent memory DIMMs (256GB per module, 1.5TB in total).
Our evaluation on Optane DC shows that its read bandwidth peaks at 37~GB/s
and the write bandwidth is 13.2~GB/s.
% and use a software layer to emulate the performance characteristics of NVM.
The server is installed with Ubuntu 19.04 and Linux kernel 5.1, the
kernel version supported by NOVA.

% \cym{To be fair, 1. How to increase thread? 2. How to allocate NVM space?}
% (2). NVDIMM has the same performance, so the collected Experimental result is valuable.

% Both
% Strata~\cite{sosp17strata} and NOVA~\cite{fast16nova} add extra NVM read latencies (300ns)
% and limit the NVM bandwidth (7.8GB/s) to emulate NVM devices. Since the PMEP hardware
% emulator~\cite{eurosys14pmfs} is unavailable,

%To evaluate the multi-core scalability, we vary the number of threads from 1 to 56 with a step of 8.
%We restrict the maximum number of threads to be 56 since we need
%4 threads to work as the \emph{master}s in KucoFS.
%We reserve the rest 4 threads for normal operating system services.

\noindent
\textbf{Compared Systems.} We evaluate KucoFS against NVM-ware file systems including
PMFS~\cite{eurosys14pmfs}, NOVA~\cite{fast16nova}, and
Strata~\cite{sosp17strata}\footnote{\url{https://github.com/NVSL/PMFS-new},
\url{https://github.com/NVSL/linux-nova}, \url{https://github.com/ut-osa/strata}},
as well as traditional file system with DAX support including Ext4-DAX~\cite{ext4dax} and XFS-DAX~\cite{xfs-dax}.
Strata only support a few applications and has trouble running
multi-threaded workloads~\cite{fast19ziggurat}, so we only give its single-threaded
performance results in Section~\ref{subsec:filebench} and Section~\ref{subsec:redis}.

Aerie is based on Linux 3.2.2, which doesn't have the
related drivers to support Optane DC.
Hence, we compare with Aerie~\cite{eurosys14aerie} by emulating persistent memory with DRAM
(\ul{Due to limited space, we only describe these experimental data in words, without 
including them in the figures}).
%Besides, it uses a hash table to
%maintain the \emph{dentry list}s so it cannot support \texttt{readdir}, and the total
%number of files in the same folder is limited by the hash table size (1~K in its source code).
%It fails to work properly with benchmarks of high concurrency.
% \begin{table}
% \centering
% \caption{Filebench workload characteristics.}
% \vspace{0.05in}
% \scalebox{0.93}[0.93]{
% \begin{tabular}{c c c c}
% \toprule[\heavyrulewidth]\toprule[\heavyrulewidth]
% \textbf{Workload} & \textbf{Avg. File Size} & \textbf{I/O Size (r/w)} & \textbf{R/W Ratio} \\
% \midrule
% Fileserver & 128~KB & 16~KB/16~KB & 1:2 \\
% Webproxy & 32~KB & 1~MB/16~KB & 5:1 \\
% Webserver & 64~KB & 1~MB/8~KB & 10:1  \\
% Varmail & 32~KB & 1~MB/16~KB & 1:1 \\
% \bottomrule[\heavyrulewidth]
% \end{tabular}}
% \label{tab:filebench}
% \vspace{-0.15in}
% \end{table}

\subsection{FxMark: Micro-benchmarks}

We use FxMark~\cite{atc16fxmark} to evaluate the basic file system
operations (in terms of both throughput and multi-core scalability). FxMark provides
19 micro-benchmarks, which
is categorized based on four criteria: data types (i.e., data or metadata), modes (i.e.,
read or write), operations (i.e., read, overwrite, append, create, etc.) and
sharing levels (i.e., low, medium or high). We only include some of them in the paper due to the limited space.

\noindent
\textbf{File Read.} Figure~\ref{fxmark-rw}~(a)-(b) show the file \texttt{read} performance of each
file system with a varying number of client threads and different sharing levels
(i.e., Low/Medium). We can observe that KucoFS exhibits significant higher
throughput than the other file systems, and its throughput scales linearly as the number of clients increases.
Specifically, with 36 client threads and \emph{Low} sharing level,
KucoFS outperforms NOVA and PMFS by 6$\times$ on average, and has two orders magnitudes 
higher performance than XFS-DAX and EXT4-DAX.
Such performance advantage stems primarily from the design of \emph{lock-free fast read},
which enables user space \emph{direct access} without the involvement of the \emph{master}. Those
kernel file systems (e.g., XFS, Ext4, NOVA and PMFS) have to perform context switch
and walk through the VFS layer, which impacts the read
performance. Besides, All of compared systems need to lock the file before actually reading the file data.
Such locking overhead impacts their performance severely, despite the contention is low~\cite{nsdi19socksdirect}.
We further observe that the throughput of the compared systems keeps steady and
low under \emph{Medium} sharing level, since all the threads are acquiring the same lock of the same file.
Instead, the performance of KucoFS is unchanged with varying sharing level, because it doesn't
rely on a per-file lock to coordinate the concurrent readers.
Note that the measured read performance via FxMark is larger than
the raw bandwidth of Optane DC (which is 37~GB/s),
because FxMark let each thread read one file page repeatedly, and the accessed data is cached in the CPU cache.
With our emulated persistent memory, Aerie shows almost the same performance
as that of KucoFS with \emph{Low} sharing level,
but its throughput becomes far behind others with \emph{Medium} sharing level. This is because
Aerie needs to contact with the TFS frequently to acquire the
lock, causing extra context switch overhead.

\begin{figure}
    \centering
    \includegraphics[width=1\linewidth]{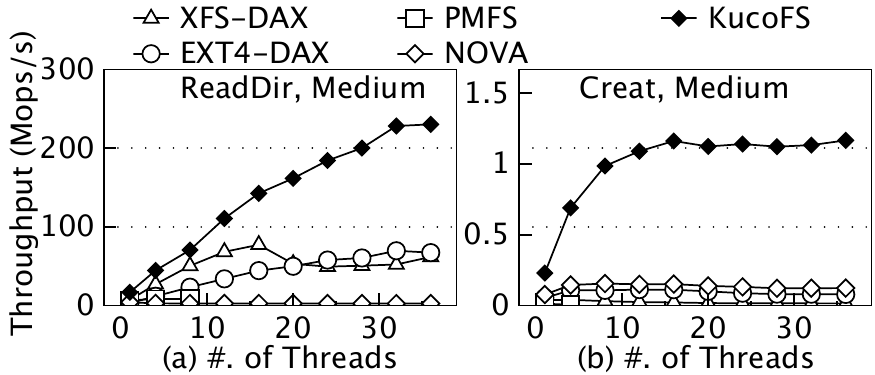}
    \vspace{-0.3in}
    \caption{\texttt{readdir} performance with FxMark. (\emph{``Medium'': In the same folder})}
    \label{fxmark-m}
    \vspace{-0.2in}
\end{figure}

\noindent
\textbf{File Write.} The throughputs of both \texttt{append}
and \texttt{overwrite} operations are given in Figure~\ref{fxmark-rw}~(c)-(e).
% We can only get the performance results of Aerie with less than 8 threads.
For overwrite operations with ``Low'' sharing level,
all systems exhibit a performance curve that increases first and then decreases.
In the increasing part, KucoFS shows the highest throughput among the compared systems because it 
is enabled to directly write data in user space. 
XFS and NOVA also shows good scalability: among them, NOVA 
partitions the free spaces to avoid global locking overhead when allocating new data pages, 
while XFS directly write data in-place without allocating new page.
Both PMFS and Ext4 fail to scale since they adopts transaction to write data, 
introducing extra locking overhead.
In the decreasing part, their throughput are restricted by the Optane bandwidth 
because of its poor scalability~\cite{izraelevitz2019basic}.
For overwrite operations with ``Medium'' sharing level, the throughput of KucoFS is
one order of magnitude higher than the other three file systems when the number of threads is small. 
Such performance benefits
mainly come from the range-lock design in KucoFS, which enables parallel updating to different data
blocks in the same file. The performance of KucoFS drops again when
the number of clients is more than 8, which is mainly restricted by the ring buffer
size in the range-lock (we reserve 8 lock items in each ring buffer).
For append operations, XFS-DAX, Ext4-DAX and PMFS exhibit un-scalable 
performance as the number of client threads
increases. This is because all of them uses a global lock to manage its metadata journal and free data pages,
so the lock contention contributes to the major overhead. Both NOVA and KucoFS show better scalability,
and KucoFS outperforms NOVA from 1.1$\times$ to 3$\times$ as the number of threads varies.
On our emulated persistent memory,
Aerie shows the worst performance because the trusted service is the bottleneck:
the clients need to frequently interact with it to acquire the lock and allocate new data pages.

We conclude that by fully exploiting the benefits of \emph{direct access}, KucoFS always shows the highest
performance among the evaluated file systems.

\noindent
\textbf{Metadata Read.}
Figure~\ref{fxmark-m}(a) shows the performance of \texttt{readdir} operations
with \emph{Medium} sharing level (i.e., all the threads read the same directory).
(Aerie doesn't support this operation).
We observe that
only KucoFS exhibits scalable performance and PMFS even cannot
complete the workloads as the number of clients increases.
These kernel file systems lock the parent directory's \emph{inode}
in VFS before reading the \emph{dentry list} and file \emph{inode}s, as a result, the execution of
different client threads is serialized when they access the same directory.
However, the skip list used in KucoFS supports lock-free reads and atomic updates, enabling multiple
readers to concurrently read the same directory.

%\begin{figure}
%    \centering
%    \includegraphics[width=1\linewidth]{expr/FxMark-MWC}
%    \vspace{-0.3in}
%    \caption{\texttt{creat} performance with FxMark. (\emph{``Low'': In different
%    folders; ``Medium'': In the same folder})}
%    \label{fxmark-mwc}
%    \vspace{-0.25in}
%\end{figure}

\noindent
\textbf{File Creation.}
To evaluate the performance of \texttt{creat} with \emph{Medium} sharing level,
FxMark lets each client thread create 10~K files in a shared directory.
% Aerie only supports 10~K files in the same folder so we only create 10~K files at most in it.
As shown in Figure~\ref{fxmark-m}(b), KucoFS achieves one order of magnitude higher throughput than
the compared file systems and it exhibits scalable performance as the number of threads increases.
XFS-DAX, Ext4-DAX and PMFS use a global lock to perform metadata journaling and manage the free spaces,
which leads to their un-scalable performance. Besides, the VFS layer needs to lock the \emph{inode} of
the parent directory before creating the files. Hence, NOVA also fails to scale despite it avoids
using global lock.
We explain the high performance of KucoFS from the following aspects:
(1) In KucoFS, all the metadata updates are delegated to the master, so it can
update them without any locking overhead. (2) By offloading
all the indexing overhead to user space, the master only needs
to do very lightweight operations. (3) KucoFS can persist
metadata with batching, while the other three kernel file
systems do not have such opportunity.
Aerie synchronizes the updated metadata of
the created files to the trusted service with batching so it achieves comparable performance as that of KucoFS, but it fails to work properly with more threads.
\begin{figure}
    \centering
    \includegraphics[width=1\linewidth]{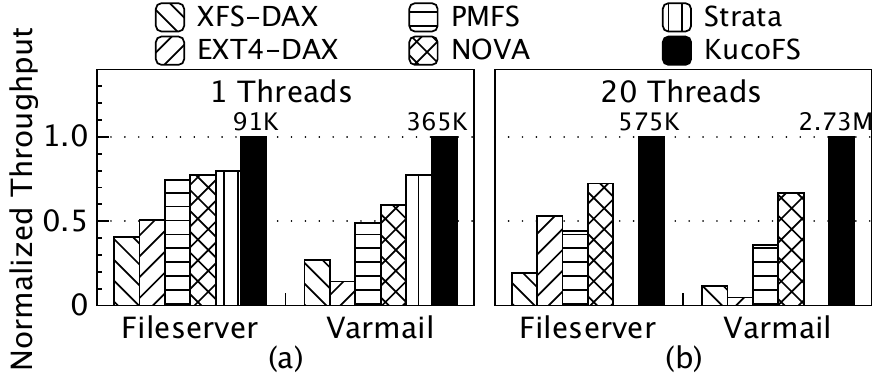}
    \vspace{-0.3in}
    \caption{Filebench Throughput with Different File Systems.}
    \label{filebench}
    \vspace{-0.2in}
\end{figure}

\subsection{Filebench: Macro-benchmarks}
\label{subsec:filebench}

We then use Filebench~\cite{filebench} as a macro-benchmark to evaluate the performance
of KucoFS. We select two workloads --- Fileserver and Varmail --- with the same
settings as that in NOVA paper: Files are created with the average size of 128~KB and 32~KB
for Fileserver and Varmail respectively. The I/O sizes of both read and write operations
are set to 16~KB in Fileserver. Varmail has read I/O size of 1~MB and write I/O size
of 16~KB. Fileserver and Varmail have write to read ratios of 2:1 and 1:1 respectively.
The total number of files in each workload is set to 100K.
Fileserver emulates I/O activity of a simple file
server~\cite{specsfs} by randomly performing creates, deletes, appends, reads
and writes. Varmail emulates an email
server and uses a write-ahead log for crash consistency. It contains a large
number of small files involving both read and write operations.
We only give single-threaded evaluation of Strata.
Figure~\ref{filebench} shows the results and we make the following observations:

(1) KucoFS shows the highest
performance among all the evaluated workloads. In single-threaded evaluation,
its throughput is 
2.5$\times$, 2$\times$, 1.34$\times$, 1.29$\times$ and 1.26$\times$ higher than XFS, Ext4,
PMFS, NOVA, and Strata respectively for Fileserver workload,
and is 2.7$\times$, 6$\times$, 2$\times$, 1.67$\times$ and 1.3$\times$ higher
for Varmail workload. Such performance advantage mainly comes from the
\emph{direct access} feature of KucoFS. It executes file I/O operations directly
in user-level, thus dismissing the OS-part overhead (i.e., context saving and reloading,
executing in VFS layer).
Strata also benefit from direct access, however, it needs to
acquire the lease from the third-party service each time they access a new file, which limits its
efficiency.
We also observe that the design of KucoFS is a good fit for Varmail workloads. This is expected:
Varmail frequently creates/deletes files, so it generates more metadata operations
and issues system calls more frequently. As described before,
KucoFS eliminates the OS-part overhead and is better at handling metadata operations.
Besides, Strata shows much higher throughput than NOVA since
the file I/Os in Varmail is small-sized. Strata only needs to append these small-sized updates
to the operation log, reducing the write amplification dramatically.

\label{subsec:redis}
\begin{figure}
    \centering
    \includegraphics[width=1\linewidth]{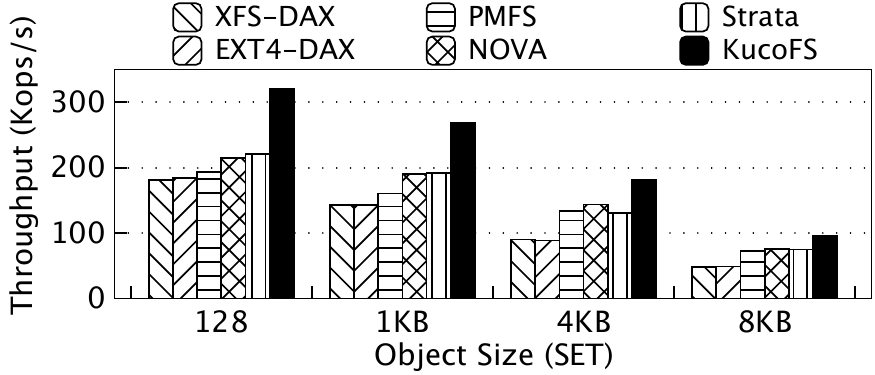}
    \vspace{-0.3in}
    \caption{Filebench Throughput with Different File Systems.}
    \label{redis}
    \vspace{-0.2in}
\end{figure}

(2) KucoFS is better at handling concurrent workloads.
With 20 concurrent client threads and Fileserver workload, KucoFS outperforms XFS-DAX and Ext4-DAX
by 3.5$\times$ on average, and PMFS by 2.3$\times$, and NOVA by 1.4$\times$.
Such performance advantage is more obvious for Varmail workload: it achieves 15\% higher performance than
XFS-DAX and Ext4-DAX on overage. Two reasons contribute to its good performance:
1) KucoFS incorporates techniques like \emph{index offloading}
to enable the \emph{master} to provide scalable metadata accessing performance;
2) KucoFS avoids using global lock by letting each client manage private free data pages.
NOVA also exhibits good scalability since it uses per-file log-structure and partitioned free space management.
%(3) To further reveal the scalability of the software design of each file system, we
%disable our NVM emulation model and measure their raw throughput directly in memory.
%As shown in Figure~\ref{filebench}~(c), the throughput of KucoFS is $1\times$ and
%$8\times$ higher than NOVA and PMFS for Fileserver workload, and 9\% and $3.5\times$
%higher for Varmail workload. As the DRAM bandwidth is no longer a constraint, the
%software overhead becomes the main part. Therefore, the advantage of \emph{direct
%access} is reflected again. We believe that these experimental results are meaningful
%since the future new devices will certainly become faster and faster.

\subsection{Redis: Real-world Application}
Many modern cloud applications use key-value stores like Redis for storing data.
Redis exports an API allowing applications to process and query structured data,
but uses the file system for persistent data storage. Redis has two approaches to
persistently record its data: one is to log operations to an append-only-file (AOF), and the
other is to use an asynchronous snapshot mechanism.
We only evaluate Redis with AOF mode in this paper.
Similar to the way in Strata~\cite{sosp17strata}, we configure Redis to use AOF mode
and to persist data synchronously.

Figure~\ref{redis} shows the throughput of SET operations using 12-byte keys and with various value sizes.
For small values, the throughput of Redis is 53\%\% higher on average on KucoFS, compared to PMFS, NOVA and Strata,
and 76\% higher compared to XFS-DAX and Ext4-DAX. This is consistent with the evaluation results of \emph{Append} operations,
where KucoFS outperforms other systems at least by 2$\times$ with a single thread.
With larger object sizes, KucoFS achieves slightly higher throughput than other file systems
since the Optane bandwidth becomes the major limiting factor.

\subsection{Benefit of Individual Optimization}
\label{subsec:opt}

\begin{figure}
    \centering
    \includegraphics[width=1\linewidth]{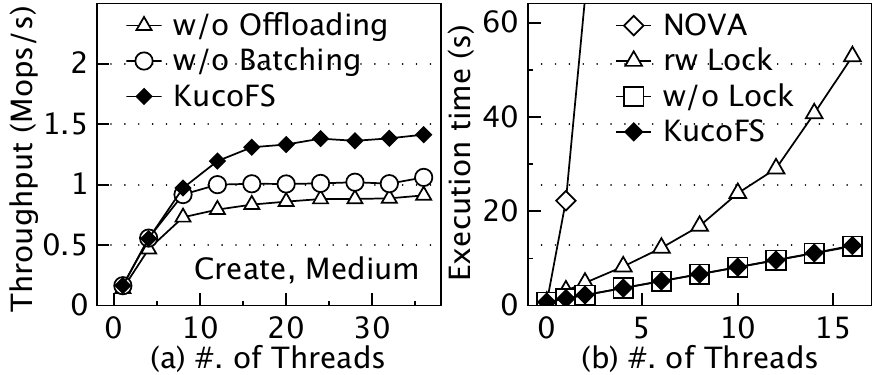}
    \vspace{-0.25in}
    \caption{Benefits of Each Optimizations in KucoFS.}
    \label{batch-offloading}
    \vspace{-0.2in}
\end{figure}

In this section, we analyze the performance improvements brought by each optimization
in KucoFS.

\ul{First, we measure the individual benefit of \emph{index offloading}
and \emph{batching-based logging}}. To achieve this, we disable batching
by letting the \emph{master} persist log entries one by one. We then move the metadata
indexing operations back to the \emph{master} to see the effects of index offloading.
Figure~\ref{batch-offloading}(a) shows the results by measuring the throughput of \texttt{creat}
with varying number of clients. We make the following observations:

%Effects of User-assisted Lookup and Batched Persisting
(1) In single thread evaluation, \emph{index offloading} does not contribute to improving
performance:  Since moving the metadata indexing from \emph{Ulib} back to the
\emph{master} doesn't reduce the total execution latency of each operation, the single-thread
throughput is unchanged. We also find that batching doesn't degrade the single-thread
performance, which is in contrast to the broad belief that batching causes higher
latency. In our implementation, the \emph{master} simply scans the message buffer
to fetch the existing requests, and the overhead of scanning is insignificant.

(2) When the number of client threads increases, we find that \emph{indexing offloading}
improves throughput by 55\% at most for \texttt{creat}
operation. Since KucoFS only allows the \emph{master} to update metadata on behalf of multiple
\emph{Ulib} instances, the theoretical throughput limit is $T_{max} = 1~req/L_{req}$
(where $L_{req}$ is the latency for a \emph{master} to process one request). Therefore, the
proposed offloading mechanism improves performance by shortening the execution time
for each request (i.e., $L_{req}$). Similarly, \emph{batching} is introduced
to speed up the processing efficiency of the \emph{master} by reducing the data persistency overhead.
From the figure, we can find that it improves throughput by 33\% at most for the \texttt{creat}
operation.

%% \vspace{15cm}
%\begin{figure}[!htb]
%    \centering
%    \includegraphics[width=1\linewidth]{expr/Lock-Free}
%    \vspace{-0.15in}
%    \caption{Effects of Lock-free Fast Read.}
%    \label{rw}
%    \vspace{-0.1in}
%\end{figure}

\ul{Second, we demonstrate the efficiency of \emph{lock-free fast read} by concurrently
reading and writing data to the same file.} In our evaluation, one read thread
is selected to sequentially read a file with I/O size of 16~KB, 
and an increasing number of threads are launched to
overwrite the same file concurrently (4~KB writes to a random offset). 
We let the read thread issues read operations for 1~million times and 
 measure its execution time by varying
the number of write threads. 
For comparison, we also implement \emph{KucoFS r/w lock}
that reads file data by acquiring the read-write lock in the range-lock ring buffer, 
and \emph{KucoFS w/o lock}
that reads file data directly without regarding the correctness.
We make the following observations from Figure~\ref{batch-offloading}(b):
(1) The proposed \emph{lock-free fast read} achieves almost the same
performance as that of \emph{KucoFS w/o lock}. This proves that the overhead
of version checking is extremely low. We also observe that
\emph{KucoFS r/w lock} needs to pay much more
time to finish reading (7\% to 3.2$\times$ more time than \emph{lock-free} for different I/O sizes).
This is because one needs to use atomic operations to acquire the range lock, and
this can severely impact read performance when there are more conflicts.
(3) The execution time of NOVA is orders of magnitudes higher than that of
KucoFS. We notice that NOVA directly uses \texttt{mutex} to synchronize the 
concurrent readers and writes. As a result, the reader will be delayed by the writers dramatically.
% \vspace{9.5cm}

\section{Related Works}
\label{sec:related}

\noindent
\textbf{Kernel/Userspace Collaboration.}
The emergence of high throughput and low latency hardware (e.g.,
Infiniband network, NVMe SSDs and NVMs) prompts the idea of moving I/O operations from the kernel
to user level: Belay et al.~\cite{osdi12dune} abstract the Dune process leveraging the
virtualization hardware in modern processors. It enables direct access to the
privileged CPU instructions in user space and executes syscalls with reduced overhead.
Based on Dune,
IX~\cite{osdi14ix} steps further to improve the performance of data-center applications
by separating management and scheduling functions of the kernel (control-plane) from
network processing (data plane).
%Arrakis~\cite{osdi14arrakis} enables the network and storage devices to
%deliver data I/Os directly to user-level applications leveraging their hardware
%virtualization capabilities, while the kernel is only involved in control-plane operations.
Arrakis~\cite{osdi14arrakis} is a new network server operating system. It splits the
traditional role of the kernel in two, where applications have direct access to virtualized
I/O devices, while the kernel only enforces coarse-grained protection and doesn't need to be
involved in every operation.
%To completely get rid of the involvements of the OS, DevFS~\cite{fast18devfs} proposes a true
%user-level direct-access file system. It achieves this by pushing the file system
%into the storage device that has compute capability and device-level RAM.
%Since the device-level processors have very limited processing capacity,
%we believe that our proposed mechanisms (e.g., load rebalance) can further improve its performance.

\noindent
\textbf{Persistent Memory File System.}
%NOVA~\cite{fast16nova}, NOVA-Fortis~\cite{sosp17nova-fortis}, PMFS~\cite{eurosys14pmfs}, SCMFS~\cite{sc11scmfs}, BPFS~\cite{sosp09bpfs}, etc.
Existing research works on NVM-based file systems can be classified into three categories:
\ding{202}\emph{Kernel-Level.}
BPFS~\cite{sosp09bpfs} adopts short-circuit shadow paging to guarantee the metadata
and data consistency. It also introduces \emph{epoch} hardware modifications to efficiently
enforce orderings. SCMFS~\cite{sc11scmfs} simplifies the file management by mapping files
to contiguous virtual address regions with the virtual memory management (VMM) in
existing OS, but it fails to support consistency for both data and metadata.
Both PMFS~\cite{eurosys14pmfs} and NOVA~\cite{fast16nova} use separated mechanisms
to guarantee the consistency of metadata and data: PMFS uses journaling for metadata updates
and perform writes with copy-on-write mechanism. NOVA is a log-structured file system deployed
on hybrid DRAM-NVM architecture. It manages the metadata with per-inode log to improve
scalability and moves file data out of the log (file data is managed with CoW) to
achieve efficient garbage collection. NOVA-Fortis~\cite{sosp17nova-fortis} steps further to be fault-tolerant
by providing a snapshot mechanism. While these kernel file systems provide POSIX I/O and propose
different approaches to enforce (meta)data consistency, their performance is still restricted
by existing OS abstraction (e.g., \emph{syscall} and VFS).
\ding{203}\emph{User-Level.} Both Aerie~\cite{eurosys14aerie} and Strata~\cite{sosp17strata}
propose to avoid the OS-part overhead by implementing the file system in user space.
With this design, user-level applications have direct access to the file system image.
Both of them adopt a third-party trusted service to coordinate the concurrent operations and
process other essential works (e.g., metadata management in Aerie and data digestion in Strata).
However, by exporting the file system image to user-level applications,
they are vulnerable to arbitrary writes from the buggy applications.
\ding{204}\emph{Device-Level.} DevFS~\cite{fast18devfs} proposes to push the file system
implementation into the storage device that has compute capability and device-level RAM, which
requires the support of dedicated hardware.
%Two concurrent works~\cite{sosp19splitfs, sosp19zofs} also aim at reducing the kernel overhead.
%Among them, SplitFS~\cite{sosp19splitfs} maps the file to user space to
%reduce kernel overhead, while still relies on an existing kernel file
%system (e.g., Ext4-DAX) to handle metadata operations; ZoFS~\cite{sosp19zofs} incorporates Memory
%Protection Keys (MPK) to enforce data protection.
%However, none of them address the protection issue caused by buggy programs.

%\noindent
%$\blacktriangleright$
%\textbf{Miscellaneous NVM Systems.}
%Except for the file systems, other existing NVM-aware systems can be roughly divided into two
%categories: One is the transactional systems, such as NV-heaps~\cite{asplos11nvheaps},
%Mnemosyne~\cite{asplos11mnemosyne}, LSNVMM~\cite{atc17lsnvmm} and DudeTM~\cite{asplos17dudetm},
%which provide transactional interfaces and uses redo/undo logging for data atomicity.
%The other is persistent data structures, such as CDDS-Tree~\cite{fast11cddstree},
%FPTree~\cite{sigmod16fptree}, NV-Tree~\cite{fast15nvtree}, Fast-Fair~\cite{fast18fastfair},
%wB+Tree~\cite{vldb15wbtree}, they provide structured data layout on a B-tree
%or its variants and have much simpler interfaces (e.g., \emph{Get/Put}).

\section{Conclusion}
\label{sec:conc}
In this paper, we revisit the file system architecture for non-volatile memories by proposing a
kernel and user-level collaborative file system named KucoFS. It fully exploits the
respective advantages of \emph{direct access} in user-level and \emph{data protection} in
kernel space. We further improve its scalability to multicores by rebalancing the
loads between kernel and user space and carefully coordinating the read and write conflicts.
Experiments show that KucoFS provides both efficient and scalable non-volatile memory management.
\newpage

%
% The next two lines define the bibliography style to be used, and the bibliography file.
\bibliographystyle{plain}
%\bibliography{ms}

\end{document}